\documentclass[%
 aip,
 sd,%
 amsmath,amssymb,
 reprint,%
]{revtex4-1}

\usepackage{graphicx}
\usepackage{dcolumn}
\usepackage{bm}
\usepackage{color}
\usepackage{braket}

\begin{document}

\preprint{AIP/123-QED}

\title[]{Decoupling bulk and surface recombination properties in silicon by depth-dependent carrier lifetime measurements}

\author{K.~Yokoyama}
\email{koji.yokoyama@stfc.ac.uk}
\affiliation{
ISIS Pulsed Neutron and Muon Facility, STFC Rutherford Appleton Laboratory, Didcot, Oxfordshire, OX11 0QX, UK
}

\author{J.~S.~Lord}
\affiliation{
ISIS Pulsed Neutron and Muon Facility, STFC Rutherford Appleton Laboratory, Didcot, Oxfordshire, OX11 0QX, UK
}

\author{J.~Miao}
\affiliation{
School of Physics and Astronomy, Queen Mary University of London, Mile End Road, London, E1 4NS, UK
}

\author{P.~Murahari}
\affiliation{
School of Physics and Astronomy, Queen Mary University of London, Mile End Road, London, E1 4NS, UK
}

\author{A.~J.~Drew}
\affiliation{
School of Physics and Astronomy, Queen Mary University of London, Mile End Road, London, E1 4NS, UK
}

\date{\today}

\begin{abstract}
Muons, as a bulk probe of materials, have been used to study the depth profile of charge carrier kinetics in Si wafers by scanning the muon implantation depth.
The photoexcited muon spin spectroscopy technique can optically generate excess carriers in semiconductor wafers, while muons can measure the excess carrier density.
As a result,  carrier recombination lifetime spectra can be obtained.
The depth-dependent lifetime spectra enable us to accurately measure the bulk carrier lifetime and surface recombination velocity by fitting the spectra to a simple 1-dimensional diffusion model.
Unlike other traditional lifetime spectroscopy techniques, the bulk and surface recombination properties can be readily de-convoluted in this method.
Here, we have applied the technique to study silicon wafers both with and without passivation treatment, and have demonstrated that the model can correctly describe the carrier kinetics in these two cases.
\end{abstract}

\maketitle

Charge carrier kinetics in semiconductors is one of the most important properties in device characterization and optimization, which ultimately determines the overall device performance.~\cite{Neamen, Schroder}
Among the various parameters that describe carrier transport, the excess carrier recombination lifetime is a key figure of merit in photovoltaics applications, which governs the efficiency of solar cells.~\cite{Steinkemper, Irvine}
In practice, the carrier lifetime of a wafer is characterized by the effective lifetime $\tau_{\rm eff}$, which can be decomposed into the bulk recombination lifetime $\tau_{b}$ and surface recombination velocity (SRV) $S$, and is often described in the approximated form: $1/\tau_{\rm eff} = 1/\tau_{b} + 2S/d$, where $d$ is the wafer thickness.~\cite{Brody}
This highlights the challenge in the recent development of high efficiency silicon solar cells ---
these new devices not only require a substrate with an extremely low impurity concentration to extend $\tau_{b}$ to the millisecond timescale, but also require excellent surface passivation in order to minimise SRV.~\cite{Yoshikawa} 
Hence it is important to characterize $\tau_{b}$ and $S$ separately, which then enables us to optimize each parameter and improve the overall device performance in a more straightforward way.
However, achieving this separation has been quite difficult because most of the existing characterization techniques (such as microwave-detected photoconductivity, photoluminescence imaging, and transient absorption/reflection spectroscopy) can measure only a bulk average, namely the effective lifetime.~\cite{Schroder}
While it is possible to decouple the bulk and surface contribution by varying the wafer thickness, this method requires multiple samples with a different $d$, which may not be easily compared due to slight variations in their surface treatment and bulk defect concentration.
Hence, there have been a number of attempts to separate the decay rates through a combination of modelling and standard measurement techniques.~\cite{Luke}
To date, however, there are still no definitive method for realizing an effective separation.

\begin{figure}
\includegraphics[width=80mm]{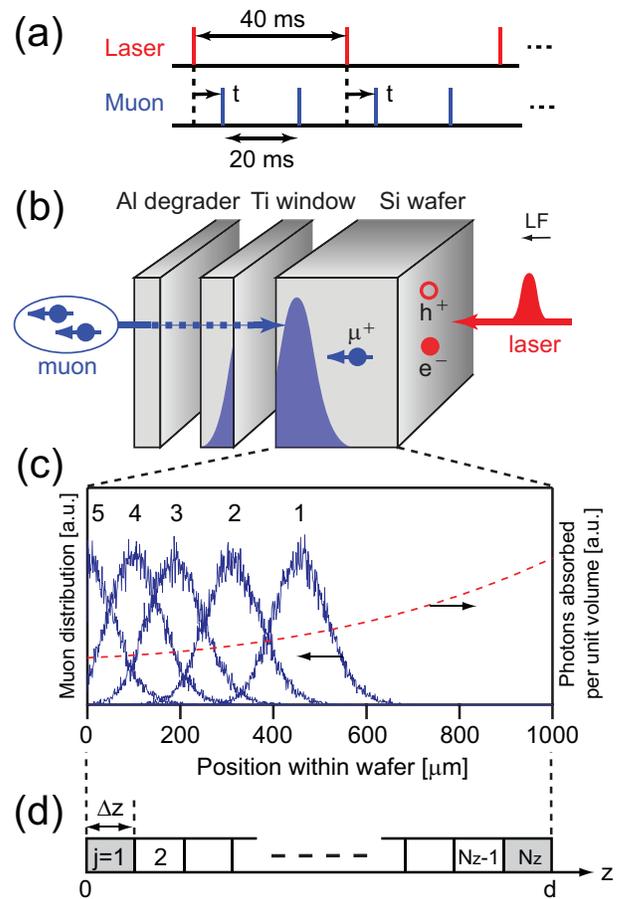}
\caption{\label{fig:Exp}
(a)
Timing diagram of the laser (15~ns FWHM pulse) and muon pulse (70~ns).
The laser pulse precedes the muon pulse by the delay t.
The muon and laser pulses operate at 50 and 25~Hz, respectively.
(b)
Schematic diagram of the experimental geometry.
Areas in blue (shadowed) illustrate the depth profile of muons distributed in both the Ti window and the sample wafer.
(c)
Calculated muon stopping distributions (solid blue lines) for five different thicknesses of the Al degrader; 0, 128, 237, 314, and 409~$\mu$m for position~1--5 respectively.
The histogram bin width is 1~$\mu$m.
An initial distribution of excess carriers at t~=~0 is plotted (red broken line) for $\alpha$~=~16~cm$^{-1}$.
(d)
Schematic diagram of the wafer segmented into N$_z$ cells ($d = N_z \Delta z$) along z-axis.
A boundary condition is applied to the end cells (filled-in with gray).
}
\end{figure}

With the aim of achieving this goal, we have developed a novel method using photoexcited muon spin relaxation~\cite{Yokoyama_RSI} (henceforth referred to as ``photo-$\mu$SR'') and have demonstrated the technique by applying it to the study of Si wafers.
This light-pump muon-probe technique utilizes the interaction between optically generated excess carriers and a hydrogen-like muonium atom Mu, which is created when an implanted positively charged muon $\mu^+$ captures an electron e$^-$ and forms a bound state.
The microscopic interaction induces spin depolarization in the initially 100~\% spin-polarized muons, and results in a relaxation of the $\mu$SR time spectra.~\cite{depolarization_mechanism}
Muon spin depolarization occurs when Mu in the triplet state, $\ket{\Uparrow \uparrow}$, where the large and small arrow denote the muon and electron spin respectively, is ionized by a free hole h$^+$ and subsequently captures a free electron.
Because the electron has its spin either in the $\ket{\uparrow}$ or $\ket{\downarrow}$ state, the regenerated Mu is in either the $\ket{\Uparrow \uparrow}$ or $\ket{\Uparrow \downarrow}$ state with equal probability, where the muon spin in the latter state will precess at the hyperfine frequency (in the order of GHz) and be depolarized.
Based on this carrier exchange interaction, one can expect quite naturally that the induced spin relaxation rate $\lambda$ will be proportional to the excess carrier density $\Delta$n.
Indeed, $\lambda$ can be used as a measure of $\Delta$n by obtaining a dependence of $\lambda$ relative to $\Delta$n, where $\Delta$n can be calculated from a measured photon fluence and attenuation coefficient $\alpha$.
This procedure, in turn, allows us to calculate $\Delta$n from a measured value of $\lambda$, and therefore the carrier lifetime spectrum (CLS), $\Delta$n(t), can be measured.
Here, t is the pulse delay between muons and pump photons, as shown in Fig. \ref{fig:Exp}(a).
Details of the procedure are described in our previous studies.~\cite{Yokoyama_PRL, Yokoyama_APL}

One of the most important advantages of this technique is that the implantation depth of muons is variable, which enables us to measure the CLS at a specified depth in a wafer.
The depth profile of implanted muons can be approximated to a Gaussian distribution with its FWHM $\approx$85~$\mu$m in the case of crystalline Si.
The muon distribution within the sample can be considered as stationary within the spatial and temporal scale of the measurement because the diffusion constants of the Mu centers are generally in the order of 10$^{-3}$~cm$^2$/s in semiconductors at RT.~\cite{Patterson}
Using a ``surface'' muon beam with an energy of 4~MeV,\cite{Blundell} the center of the muon distribution can be scanned between the sample surface and the maximum implantation depth, which is $\approx$700~$\mu$m in Si.
The muon stopping position can be adjusted by placing material (called a ``degrader'') in the muon beam before the sample to modify the incident beam energy.
Hence, the CLS can be measured at several depths, which provides a sterical picture of the excess carrier kinetics.
The set of lifetime spectra can be fitted simultaneously to a 1-dimensional (1-D) diffusion model, which numerically calculates the dynamics of excess carriers with $\tau_{b}$ and $S$ as fit parameters.
In this way, the recombination parameters can be measured separately with good accuracy.

To demonstrate the method, a photo-$\mu$SR experiment studying Si wafers was carried out using the HiFi spectrometer at the ISIS Neutron and Muon Source at the STFC Rutherford Appleton Laboratory in the UK.~\cite{Lord}
A detailed discussion of this unique laser facility and associated equipment can be found elsewhere.~\cite{Yokoyama_RSI, Yokoyama_PRL}
In this experiment, 100\% spin-polarized muons are incident on one side of the sample, while pump light illuminates the other side, as shown in Fig. \ref{fig:Exp}(b).
Here, the sample is a 2-inch diameter, 1~mm thick, float-zone grown single crystal Si wafer (intrinsic, R$>$10000 $\Omega$cm, both sides chemically polished) with the $<$111$>$ axis perpendicular to the surface.
The sample was contained in a helium-gas purged cell, with a 100~$\mu$m thick titanium window on one side (for muon entry) and a glass window to admit light on the other side. 
The gas purged construction ensured that the sample temperature was kept at 290~K throughout the experiment.
A small magnetic field (100~G) was applied parallel to the initial $\mu^+$ spin direction (longitudinal field, LF) for decoupling the intrinsic relaxation in ``dark'' $\mu$SR spectra.
This relaxation may be due to trace impurities in the sample or field inhomogeniety of the instrument,~\cite{Yokoyama_PRL} while nuclear moments of Si$^{29}$, an isotope with spin $\frac{1}{2}$, magnetic moment of -0.555 (in nuclear magnetons), and 4.7~\% natural abundance, can also contribute.
Aluminum foils were used as a degrader for decelerating muons and adjusting the implantation depth.
Note that muons stopped in the Al degrader or Ti window (especially for position~5) form diamagnetic centers and only give a flat background in the observed $\mu$SR spectra.
This fraction can be ignored in the analysis because $\Delta$n is determined solely by the relaxation rate.
As shown in Fig. \ref{fig:Exp}(c), the implantation depth was scanned from position 1 to 5 by increasing the thickness of the Al degrader.
The spatial distribution for each case was calculated with the help of ``musrSim''~\cite{Sedlak} (a Monte Carlo simulation package based on GEANT4) using the known incoming muon momentum along with the amount and areal density of materials in the beam.
The Nd:YAG fundamental beam was generated and used as a pump laser light.~\cite{Yokoyama_RSI}
The wavelength, 1064~nm, typically gives $\alpha$~$\approx$~10~cm$^{-1}$ in intrinsic Si at RT, which generates excess carriers throughout the 1~mm thick wafer.
Multiple (partial) reflections from the wafer surfaces and windows were taken into account in the calculation.
The spatial profile of laser light was nearly uniform on the sample, and large enough to cover the entire muon spot --- this condition allows us to assume a laterally uniform system which simplifies the model.

Two identical wafers with different surface treatments were measured.
Sample~A is an as-is wafer, with the only preparation carried out being solvent cleaning.
In this case the wafer was immersed in the warm acetone bath, followed by rinsing with methanol and deionized water;
in this process, oil particles and organic residues on the Si surface are loosened and removed.
The wafer was then dried with nitrogen gas and loaded into the sample cell under a high-purity He-gas atmosphere.
In contrast, Sample~B was cleaned with two additional steps to passivate the surface.~\cite{Tian}
In this case, after the preparatory solvent clean, the procedure involved a standard RCA clean in an aqueous solution of ammonium hydroxide and hydrogen peroxide, where the hot solution (heated to $\approx$70~$^\circ\mathrm{C}$) removed remaining hydrocarbon residues, oxidized the surface and formed a thin oxide layer.
The wafer was then immersed in dilute hydrofluoric (HF) acid (2~\%) for approximately three minutes.
The HF acid etches the silicon dioxide layer and produces a clean H-terminated $<$111$>$ surface, whose hydrophobicity can be checked with a wetting test.
It is known that this method works particularly well for capping dangling bonds on Si $<$111$>$ surfaces by forming covalent Si-H bonds.
As a result, wafers prepared in this manner have exceptionary low SRV.~\cite{Yablonovitch, Grant}
After that, the wafer was taken out to ambient atmosphere and moved into the He-gas atmosphere.
During sample loading, care was taken to minimize exposure time to atmosphere when there could be a gradual deterioration of the H-terminated surfaces due to oxidation.
Once the wafer was sealed in the purged cell, the oxidation rate should be very slow.

\begin{figure}
\includegraphics[width=80mm]{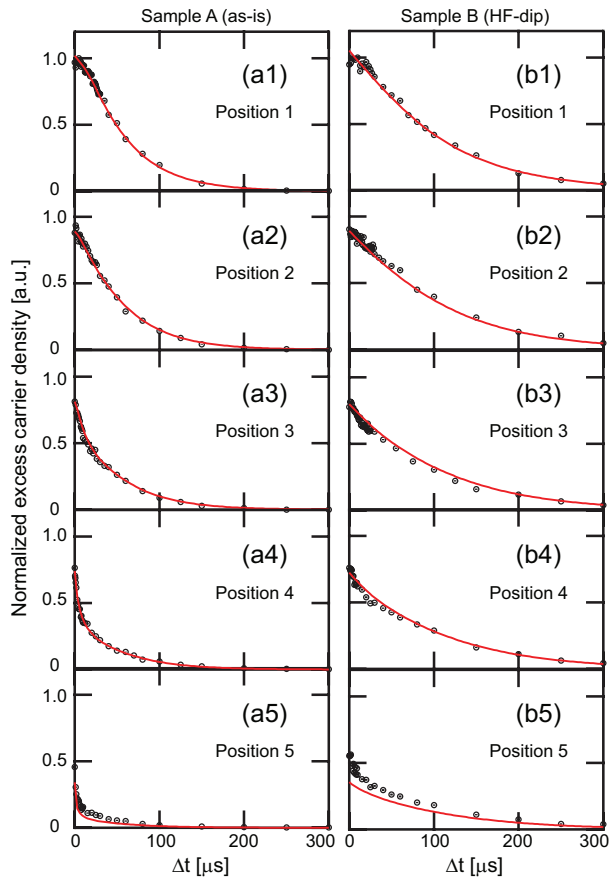}
\caption{\label{fig:LTS}
(a1) -- (a5) and (b1) -- (b5): Carrier lifetime spectra measured at position 1-5 for Sample A and B, respectively.
Each set is normalized to $\Delta$n(t = 0) at position 1, which is 1.2~$\times$~10$^{14}$ and 1.0~$\times$~10$^{14}$ cm$^{-3}$ for Sample A and B, respectively.
Red solid lines denote the best fit from the modeling described in the text.
}
\end{figure}

Fig. \ref{fig:LTS}(a1)-(a5) show a sequence of CLS measured for Sample A at the five depths indicated in Fig. \ref{fig:Exp}(b).
The spectra exhibit characteristic excess carrier kinetics with a fast recombination taking place in the range closer to the surface.
Generated free electrons and holes diffuse together at RT because of the Coulomb interaction (ambipolar diffusion).~\cite{Yokoyama_APL}
Based on the assumption of lateral uniformity, their motion can be described by the 1-D diffusion equation for $\Delta n(z,t)$,

\begin{eqnarray}
\frac{\partial \Delta n}{\partial t} = D\frac{\partial^{2}\Delta n}{\partial z^{2}}-\frac{\Delta n}{\tau_{b}}
\;
\label{eq:bulk},
\end{eqnarray}

\noindent
where $D$ is the ambipolar diffusion constant, and $z$ is the depth within the sample along the axis of the muon and laser beams.
The surface of the wafer on which the muons are incident is set as z~=~0 [see Fig. \ref{fig:Exp}(d)].
The laser pulse excites a carrier distribution in the wafer instantaneously because its temporal duration, $\approx$15~ns FWHM, is much shorter than the timescales of the carrier dynamics.
The initial carrier distribution, as shown in Fig. \ref{fig:Exp}(c), can be calculated based on the photon flux on the sample, attenuation coefficient, and reflection coefficients of the sample and window surfaces.
To solve this diffusive initial value problem, Eq. \ref{eq:bulk} is numerically integrated using the Lax-Friedrichs method~\cite{Press}, with a finite time step $\Delta$t and the wafer segmented into cells of width $\Delta$z [see Fig. \ref{fig:Exp}(d)].
The algorithm gives a stable solution when the stability criterion, $2 D \Delta t / (\Delta x)^2 \leq 1$, is satisfied.
Clearly, the calculation will require more computational resources as the timescale of the system becomes longer.
With a view to applying this technique to other systems with a longer carrier lifetime, many of the calculations in our code are implemented as array operations for speed.~\cite{SM}
For this study, the numerical calculations were carried out with $\Delta z$ = 5~$\mu$m and $\Delta t$ = 1~ns, which would provide sufficient spatial resolution for weighting the series of CLS with the muon distribution (see below), while finishing the integration within reasonable amount of time.
The calculation then gives us a CLS for each cell in Fig. \ref{fig:Exp}(d) {\it i.e.} $\Delta n_j(t)$, where $j = [1, N_z]$.
In order to fit it against the measurement, we need to take into account the spatial distribution of muons by weighting $\Delta n_j(t)$ with the normalized distribution histogram $H_j$, before summing over the entire range to obtain a lifetime spectrum $\Delta n^p(t)$, {\it i.e.} $\Delta n^p(t) = c \sum_{j = 1}^{N_z} H^p_{j} \Delta n_j(t)$, where the superscript $p$ denotes the positions as defined in Fig. \ref{fig:Exp}(c), and $c$ is a scaling factor.
Here, $H^p_j$ is defined by $H^p_j = h^p_j / \sum_{j = 1}^{N_z} h^1_j$, where $h^p_j$ is the number of muons implanted in the $j$-th cell for the measurement carried out in position $p$.
Note that $\sum_{j = 1}^{N_z} h^1_j$ essentially gives the total number of muons incident on the sample without the degraders.
Therefore, we can scale each $\Delta n^p(t)$ to the corresponding data with the single scaling factor.

For Sample~A, we can assume a Dirichlet boundary condition, where $\Delta n(z=0,t) = \Delta n(z=d,t) = 0$, because the SRV for unpassivated wafers approaches the thermal velocity and hence both surfaces essentially act as an infinite sink for excess carriers.
Simultaneous curve fitting was performed for the five lifetime spectra measured, with $D$, $\tau_b$, $\alpha$, and $c$ as fit parameters.
Calculated curves for the best fit to the data are shown in Fig. \ref{fig:LTS}(a1)-(a5), with the fit parameters summarized in Table \ref{table:FitResults}.
The fit curves are in a good agreement with the measurement, indicating that the numerical model is able to describe the carrier dynamics at all depths in the sample.

\begin{table}
\centering
\caption{Summary of fit results}
\label{table:FitResults}
\begin{tabular}{p{10mm}>{\centering\arraybackslash}p{13mm}>{\centering\arraybackslash}p{13mm}>{\centering\arraybackslash}p{13mm}>{\centering\arraybackslash}p{13mm}>{\centering\arraybackslash}p{13mm}}
\hline
\hline
Sample & D & $\tau_b$ & S & $\alpha$ & c\\
 & [cm$^2$/s] & [$\mu$s] & [cm/s] & [1/cm]\\
\hline
A & 10.5(2) & 100(2) & --- & 13.9(2) & 0.1326(2)\\
B & 7.1(3) & 107(1) & 106(6) & 15.9(1) & 0.1364(4)\\
\hline
\hline
\end{tabular}
\end{table}

The lifetime spectra for Sample~B were measured in the same manner and results are shown in Fig. \ref{fig:LTS}(b1)-(b5).
In contrast to Sample~A, the decay rates are significantly slower for all depths measured due to the slower surface recombination in this sample.
In order to characterize this behavior, the fit model now applies a finite SRV to the end cells, such that $\Delta n_{1}$ and $\Delta n_{N_z}$ decrease by $e^{-S \Delta t/\Delta z}$ at every time step.
The same fitting method was applied to the lifetime spectra, with $S$ as an additional fitting parameter.
Again, the fits agree well with the measurements, indicating that the model can describe the carrier behavior with both bulk and surface recombination dynamics.
As shown in Table \ref{table:FitResults}, the fit parameters, $D$, $\tau_b$, and $\alpha$, agree with the results from Sample~A.
This is expected because the wafers are cut from the same single crystal ingot, and these bulk properties should be in the same range.
The measured SRV, 106~$\pm$~6~cm/s, is significantly slower than the thermal velocity of carriers in Si ($\approx$10$^7$~cm/s), but is faster than the reported values for samples immersed in HF acid ($\approx$1~cm/s).~\cite{Grant}
In our setup, it was inevitable that the passivated surfaces on Sample~B was exposed to air and slightly deteriorated during the transfer from the HF bath to the purged cell.
However, the surface treatment slowed down the SRV sufficiently to allow us to demonstrate the method.

While the fit curves in Fig. \ref{fig:LTS} generally agree well with the data, small discrepancies can be observed in Fig. \ref{fig:LTS}(a5) and (b5).
The reason for this is that the spatial distribution of $\Delta$n near the surface is highly non-uniform due to the fast SRV.
As previously discussed, the method determines $\Delta$n from the muon spin relaxation rate, which has been averaged over the muon distribution width.
If the distribution is centered around the surface, where $\Delta$n(z) changes significantly over the muon range, the method may not have sufficient spatial resolution to characterize the positional dependence.
In addition, the distribution calculation with musrSim assumed exact beam properties, which might be slightly different from the actual beam during the experiment.
This will matter most when close to the surface with large values of $d\Delta n$/$dz$.
To resolve these problems, one can simply avoid measuring the surface region, instead increasing the number of depth measurements sampled in the bulk.
This should still allow a sufficient number of lifetime spectra to be obtained to allow the parameters determined with good accuracy.

In conclusion, we have demonstrated that photo-$\mu$SR is capable of measuring the CLS at multiple depths in a Si wafer by changing the implantation depth of a muon beam.
The depth-dependent lifetime spectra enable us to accurately measure the bulk carrier lifetime and SRV by fitting the spectra to a simple 1-dimensional diffusion model.
Unlike other traditional lifetime spectroscopy techniques, the bulk and surface recombination properties can effectively be decoupled using this method, with values independently determined for each.
In addition to providing new insights into surface recombination, future work may allow us to determine the intrinsic lifetime limit for Si, which essentially defines the maximum efficiency of Si solar cells.
Utilizing the penetrating power of surface muons, the technique can be applied to a complete solar cell to measure carrier lifetimes, thus completing a measurement that is difficult to make with standard methods.
Furthermore, these measurements may benefit from temperature-dependent studies ({\it e.g.} to better understand the properties of solar cells used in extreme conditions).
These type of measurements would be straightforward with the He-purged sample cell and existing temperature control setup.
It is also worth noting that the method should be applicable to other semiconductor systems, as long as the bulk carrier lifetime is sufficiently long (at least a few $\mu$s) so that the $\Delta$n(z) does not change significantly during the $\mu$SR time window used to determine the relaxation rate.~\cite{Yokoyama_PRL, Yokoyama_APL}
Therefore, we anticipate a wide range of future studies, including active semiconductor materials for power electronics devices, such as silicon carbide, where the carrier lifetime directly relates to the device switching efficiency.

This work was carried out using a beamtime allocated by the STFC ISIS Facility.~\cite{DOI}
Additionally, part of this work was supported by European Research Council (Proposal No 307593 -- MuSES).
We would like to thank Dr. Stephen Cottrell of the ISIS Muon group for providing us valuable suggestions and discussions.
Finally, we are grateful for the assistance of a number of technical and support staff in the ISIS facility.



\begin{thebibliography}{10}
\bibitem{depolarization_mechanism} The depolarization mechanism described here is a simplified model.
In reality, it involves Mu states in different charge states located at multiple crystalline sites, which form a complex network of dynamic transitions.
For further information on the depolarization mechanism in Si, see, for example, Ref. \cite{Yokoyama_PRL} and references therein.
\end{thebibliography}
\end{document}